\documentclass{article}
\usepackage{graphicx}
\def\no{\noindent}
\def\bc{\begin{center}}
\def\ec{\end{center}}
\def\vs{\vskip0.5cm}
\def\beq{\begin{equation}}
\def\eeq{\end{equation}}

\def\d{\downarrow}
\def\u{\uparrow}

\def\bj{{\bf j}}

\begin{document}

\title{Jahn-Teller systems at half filling: \\
crossover from Heisenberg to Ising behavior
}

\author{K. Ziegler \\
Institut f\"ur Physik, Universit\"at Augsburg, D-86135 Augsburg, Germany \\
}

\maketitle

\no
Abstract:

\no
The Jahn-Teller model with $E\otimes\beta$ electron-phonon coupling and 
local (Hubbard-like) Coulomb interaction is considered 
to describe a lattice system with two orbitals per site at half filling.
Starting from a state with one electron per site, we follow the tunneling
of the electrons and the associated creation of an 
arbitrary number of phonons due to electron-phonon interaction.
For this purpose
we apply a recursive method which allows us to organize systematically
the number of pairs of empty/doubly occupied sites and to include 
infinitely many phonons which are induced by electronic tunneling. In 
lowest order of the recursion (i.e. for all processes with only 
one pairs of empty/doubly occupied sites)
we obtain an effective anisotropic pseudospin 1/2 Heisenberg 
Hamiltonian $H_{eff}$ as a description of the orbital degrees of freedom.
The pseudospin coupling depends on the physical parameters and the energy. 
This implies that the resulting resolvent $(z-H_{eff}(z))^{-1}$
has an infinite number of poles, even for a single site. 
$H_{eff}$ is subject to a crossover from an isotropic
Heisenberg model (weak electron-phonon coupling) to an Ising model (strong 
electron-phonon coupling).

\maketitle


\section{Introduction}

It has been known for a long time that orbital degrees of freedom in
a systems with Jahn-Teller coupling can
be described by an effective pseudospin Hamiltonian 
\cite{kugel73,feiner97,harris04}. The coupling parameter of the pseudospin
interaction is $t^2/U$, where $t$ is the orbital hopping rate and
$U$ is the strength of the on-site Coulomb interaction. This
description is of great interest because it provides a model
to study orbital ordering and the possibility of orbital 
liquids in terms of conventional spin theories.

In this paper the influence of the electron-phonon 
coupling strength on the effective pseudospin Hamiltonian will studied
in detail. In order to keep the calculations simple only the case of a
system with $E\otimes\beta$ Jahn-Teller coupling is considered, and
the electron spin is neglected.
A recursive projection formalism \cite{ziegler03} is applied to 
derive the effective pseudospin Hamiltonian. This approach provides
pseudospin coupling parameters that depend on the electron-phonon 
coupling strength. 

The paper is organized as follows. In Sect. 2 the model is defined.
As a physical quantity the resolvent, related to the electron-phonon
Hamiltonian, is considered. Its relation with physical quantities is
discussed in Sect. 2.1. The recursive projection method is briefly
described in Sect. 3 and the effective pseudospin Hamiltonian, obtained 
from this method, is presented in Sect. 4. Finally, the crossover from
weak to strong electron-phonon coupling is studied in Sect. 4.1.
 
\section{The $E\otimes\beta$ Jahn-Teller Model}

The Jahn-Teller model describes fermions with pseudospin $\sigma=\u ,\d$,
coupled to phonons. It is defined by the Hamiltonian $H=H_t+H_0$,
where $H_t$ is the hopping term of the fermions between nearest-neighbor
sites $\bj$ and $\bj'$
\[
H_t=-t\sum_{<\bj,\bj'>}\sum_{\sigma=\u,\d}
c_{\bj\sigma}^\dagger c_{\bj'\sigma}+h.c.
\]
and $H_0$ is a local (Hubbard-like) interaction and a phonon term:
\[
H_0=\sum_{\bj}\Big[\omega_0 b_\bj^\dagger b_\bj
+g(b_\bj^\dagger+b_\bj)(n_{\bj\u}-n_{\bj\d})
+Un_{\bj\u}n_{\bj\d}\Big]
\]
for dispersionless phonons with energy $\omega_0$.
For a given ensemble of fermions, represented by integer
numbers $n_{\bj \sigma}=0,1$, the Hamiltonian $H_0$
can be diagonalized with product states
\[
\prod_\bj |N_\bj,n_{\bj\u},n_{\bj\d}\rangle,
\]
where $N_\bj =0,1,...$ is the number of phonons at site $\bj$.
The state $|N_\bj,n_{\bj\u},n_{\bj\d}\rangle$ is an eigenstate of 
the phonon-number operator $b_\bj^\dagger b_\bj$ for
$n_{\bj\u}=n_{\bj\d}=0$ and for $n_{\bj\u}=n_{\bj\d}=1$:
\[
b_\bj^\dagger b_\bj|N_{\bj},0,0\rangle
=N_\bj|N_{\bj},0,0\rangle\ \ \ 
b_\bj^\dagger b_\bj|N_{\bj},1,1\rangle 
=N_\bj|N_{\bj},1,1\rangle.
\]
The corresponding eigenstates with a single fermion at $\bj$ are
obtained from $|N_\bj,0,0\rangle$ as
\beq
|N_\bj,1,0\rangle=c_{{\bf j}\u}^\dagger\exp[-{g\over\omega_0}(b_\bj^\dagger
-b_\bj)]|N_\bj,0,0\rangle,\ \ \ 
|N_\bj,0,1\rangle
=c_{{\bf j}\d}^\dagger\exp[{g\over\omega_0}(b_\bj^\dagger
-b_\bj)]|N_\bj,0,0\rangle
\label{basis}
\eeq
and
\[
|N_\bj,1,1\rangle=c_{{\bf j}\u}^\dagger
c_{{\bf j}\d}^\dagger|N_\bj,0,0\rangle.
\]
$\prod_\bj |N_\bj,n_{\bj\u},n_{\bj\d}\rangle$ is an eigenstate of $H_0$
with energies
\beq
E_0(\{N_\bj,n_{\bj\u},n_{\bj\d}\})
=\sum_{\bj}\Big[\omega_0 N_\bj-{g^2\over\omega_0}(n_{\bj\u}-n_{\bj\d})^2
+Un_{\bj\u}n_{\bj\d}\Big].
\label{energy0}
\eeq
The groundstate of $H_0$ has no phonons.

\subsection{The Resolvent}

In the following the resolvent
\[
(iz+H)^{-1}
\]
shall be studied. It is directly related to a number of physical
quantities. One is linked with the thermodynamic properties
of a statistical ensemble governed by the Hamiltonian $H$ through
the Boltzmann weight at inverse temperature $\beta$. It reads
\beq
e^{-\beta H}=\int_\Gamma (iz+H)^{-1}e^{i\beta z}{dz\over2\pi}.
\label{resolvent0}
\eeq
$\Gamma$ is a closed contour that encloses all eigenvalues of $H$.
Another connection is with the dynamics of quantum states in a system
characterized by the Hamiltonian $H$: The evolution of a state
$|\Psi_0\rangle$ at time 0 to a later time $t$ is given by
\[
|\Psi_t\rangle =e^{iHt}|\Psi_0\rangle .
\]
A Laplace transformation for positive time gives with $Im z<0$
the resolvent that acts on the initial state:
\beq
\int_0^\infty e^{-izt}|\Psi_t\rangle dt
=\int_0^\infty e^{-izt}e^{iHt}dt |\Psi_0\rangle
=(z-H)^{-1}|\Psi_0\rangle.
\label{evol1}
\eeq
The return probability to the initial state is obtained from
the inverse Laplace transform and reads
\beq
\langle\Psi_0|\Psi_t\rangle={1\over2\pi}\int_{-\infty}^\infty e^{izt}
\langle\Psi_0|(z-H)^{-1}|\Psi_0\rangle dz.
\label{integr0}
\eeq
Using the spectral representation of $H$ with eigenvalues $E_j$,
the expectation value in the integrand reads
\[
\langle\Psi_0|(z-H)^{-1}|\Psi_0\rangle = 
\sum_j{|\langle E_j|\Psi_0\rangle|^2\over z-E_j}.
\]
Inserting this into Eq. (\ref{integr0}) allows us to
apply Cauchy's Theorem to perform the integration. 

In order to evaluate the resolvent a standard procedure is to 
expanded it in terms of $H_t$ in a Neumann series
\[
(z-H)^{-1}=(z-H_0-H_t)^{-1}=(z-H_0)^{-1}\sum_{l\ge0}[H_t(z-H_0)^{-1}]^l
\] 
and to truncate this series. The poles of any finite truncation are 
the eigenvalues of the unperturbed Hamiltonian $H_0$. This may be 
insufficient for a good
approximation of $(z-H)^{-1}$. In the next section a recursive approach 
is applied that avoids this restriction. 

\section{Projection Formalism and Continued-Fraction Representation}

Considering a half-filled system, the projected resolvent 
$P_0(z-H)^{-1}P_0$ must be evaluated. It is assumed that
$P_0$ projects the states of the entire Hilbert ${\cal H}$
space to the subspace ${\cal H}_0$.  The projected resolvent satisfies
the identity
\beq
P_0(z-H)^{-1}P_0=\Big[
P_0(z-H)P_0 - P_0H P_1 (z-H)_1^{-1}P_1 HP_0\Big]_0^{-1},
\label{projected}
\eeq
where $P_1={\bf 1}-P_0$ projects onto the Hilbert space ${\cal H}_1$ that is
complementary to ${\cal H}_0$. If $H$ obeys the relations
\[
P_0HP_1=P_0HP_2,\ \  P_1HP_0=P_2HP_0\ \ \ (P_2\ne P_1),
\]
Eq. (\ref{projected}) can also be written as
\beq
P_0(z-H)^{-1}P_0=\Big[P_0(z-H)P_0 - P_0H P_2 (z-H)_1^{-1}P_2 HP_0\Big]_0^{-1}.
\label{projected2}
\eeq
The identity used in Eq. (\ref{projected}) can be applied again to
$P_2 (z-H)_1^{-1}P_2$ on the right-hand side. This procedure can be
applied iteratively. It creates a hierarchy of projectors $P_k$ 
onto Hilbert spaces ${\cal H}_k$. It is based 
on the fact that the projector $P_{2j+1}$ 
is created from $P_{2j-1}$ and $P_{2j}$ as
\[
{\cal H}_{2j+1}={\cal H}_{2j-1}\backslash {\cal H}_{2j}
\subset {\cal H}_{2j-1},
\]
and $P_{2j+2}$ comes from the relation
\beq
P_{2j+1}HP_{2j}=P_{2j+2}HP_{2j}\equiv H_{j+1,j}
\ \ \ {\rm and}\ \ \ 
P_{2j}HP_{2j+1}=P_{2j}HP_{2j+2}\equiv H_{j,j+1}.
\label{project2}
\eeq
In terms of the projected resolvents this construction implies  
a recursion relation.
Using $G_{2j}=P_{2j}(z-H)_{2j-1}^{-1}P_{2j}$ and $H_{j,j+1}=P_{2j}HP_{2j+2}$
this reads
\beq
G_{2j}=\Big[z-P_{2j}HP_{2j} 
- H_{j,j+1}G_{2j+2}H_{j+1,j}\Big]_{2j}^{-1}.
\label{projected3}
\eeq
It is useful that the Hamiltonian $H$ in Sect. 2 is the sum of
two Hamiltonians $H=H_0+H_1$ and that one can choose projections such 
that the following holds: 

\no
(1) $H_0$ stays inside the projected Hilbert space: $H_0P_0=P_0H_0P_0$
and $P_0H_0=P_0H_0P_0$.

\no
(2) $H_1$ maps from ${\cal H}_{2j}$ to ${\cal H}_{2j+2}$:
\[
H_1: {\cal H}_{2j}\to {\cal H}_{2j+2},
\]
where ${\cal H}_{2j}$ is orthogonal to ${\cal H}_{2j+2}$. In the next
section these properties will be used to construct an effective 
Hamiltonian.

\section{The Effective Hamiltonian}

The projection formalism is now applied to the Hamiltonian $H=H_0+H_t$
of Sect. 2. The Hilbert space separates into subspaces with a fixed
number of fermions with pseudospin $\u$ and a fixed number with 
pseudospin $\d$ because $H$ cannot change it with a diagonal pseudospin
term. The case is considered here where $P_0$ projects
onto singly occupied sites with no phonons. $H_t$ is off-diagonal with
respect to the phonons and changes
the number of pairs of empty/doubly-occupied sites (PEDS) by one. 
Therefore, $P_{2j}$ ($j\ge1$) projects onto states with $j$ PEDS and any
number of phonons. According to this construction, the matrix elements of
$P_{2j}HP_{2j+2}=P_{2j}H_tP_{2j+2}$ and $P_{2j+2}HP_{2j}=P_{2j+2}H_tP_{2j}$
are all non-zero with respect to different phonon numbers.
$P_{2j}H_0P_{2j}$ is diagonal in the basis (\ref{basis}) with
matrix elements:
\[
\omega_0\sum_{\bj}N_\bj-2(M-j)g^2/\omega_0+Uj,
\]
where $2(M-j)$ counts the number of singly-occupied sites and $j$ the number
of doubly-occupied sites on a lattice
with $2M$ sites. Thus the recursion relation of Eq. (\ref{projected2}) 
becomes
\beq
G_{2j}=\Big[z-\omega_0\sum_{\bj}N_\bj+ 2(M-j)2g^2/\omega_0-Uj 
- H_{j,j+1}G_{2j+2}H_{j,j+1}^T\Big]_{2j}^{-1}
\label{projected4}
\eeq
with $H_{j,j+1}$ defined in Eq. (\ref{project2}).
The recursion terminates on a finite lattice ($M<\infty$) if $j=M$,
since at most $M$ PEDS can be created and, therefore, $P_{2M+1}$ is 
a projection onto the empty space. $G_{2M}$ is diagonal with
matrix elements
\[
{1\over z-\omega_0\sum_{\bj}N_\bj-UM}.
\]
This can serve as a starting point for the iterative approximation of
$G_0$. On the other hand, we can express $G_0$ by $G_2$ through 
Eq. (\ref{projected4}) and {\it approximate} $G_2$ by the diagonal matrix
\beq
G_2\approx [z-\omega_0\sum_{\bj}b^\dagger_\bj b_\bj
+ 2(M-1)g^2/\omega_0-U ]_{2}^{-1}.
\label{g2}
\eeq
This approximation corresponds with a truncation of all scattering 
processes with more than one PEDS. It is valid for weak
hopping, i.e. for $t/\omega_0$ small in comparison with $g/\omega_0$
and $U/\omega_0$. Subsequently, it will turn out that an effective 
expansion parameter for the truncation is 
$t^2/(U/\omega_0+2g^2/\omega_0^2)$.
  
Eq. (\ref{g2}) gives for $G_0$ the expression
\beq
G_{0}=\Big[z+ 2Mg^2/\omega_0
- H_{0,1}[z-\omega_0\sum_{\bj}b^\dagger_\bj b_\bj
+ 2(M-1)g^2/\omega_0-U ]_{2}^{-1}H_{0,1}^T\Big]_{0}^{-1}.
\label{recurse2}
\eeq
$H_{0,1}[...]_{2}^{-1}H_{0,1}^T$
is a matrix in a Hilbert space without phonons according to the
definition of $P_0$. $H_{0,1}^T$ creates phonons as well as a PEDS.
This will be picked up by the diagonal matrix $[...]_2^{-1}$. Finally,
$H_{0,1}$ annihilates the phonons and the PEDS. Consequently, the
entire expression is either diagonal or has off-diagonal elements
with nearest-neigbor pseudospin exchanges. Such a matrix can be expressed
by a (anisotropic) pseudospin-1/2 Hamiltonian. A detailed calculation 
gives an anisotropic Heisenberg Hamiltonian 
\[
H_{eff}\equiv
-2Mg^2/\omega_0
+ H_{0,1}[z-\omega_0\sum_{\bj}b^\dagger_\bj b_\bj
+ 2(M-1)g^2/\omega_0-U ]_{2}^{-1}H_{0,1}^T
\]
\beq
=-2Mg^2/\omega_0+\sum_{{\bf j},{\bf j}'}\Big[
a_{\u\u}(S_{\bf j}^zS_{{\bf j}'}^z-{1\over4})+
a_{\u\d}(S_{\bf j}^xS_{{\bf j}'}^x+S_{\bf j}^yS_{{\bf j}'}^y)
\Big]
\label{heff}
\eeq
with the pseudospin-1/2 operators $S^x,S^y,S^z$ and $z$-dependent 
coupling coefficients (cf. Appendix A)
\[
a_{\u\u}=2t^2 {e^{-2g^2/\omega_0^2}\over\omega_0}
\gamma^*({U-z-2(M-1)g^2/\omega_0\over\omega_0},-2g^2/\omega_0^2)
\]
\[
a_{\u\d}=2t^2 {e^{-2g^2/\omega_0^2}\over\omega_0}
\gamma^*({U-z-2(M-1)g^2/\omega_0)\over\omega_0},2g^2/\omega_0^2).
\]
$\gamma^*$ is the incomplete Gamma function \cite{abramowitz}:
\[
\gamma^*(a,y)=\sum_{m\ge0}{1\over m!}{(-y)^m\over a+m}.
\]

Introducing the system energy $E=z+2Mg^2/\omega_0$ we can write
$
(z-H_{eff})^{-1}=(E-H_H)^{-1},
$
where $H_H$ is the Heisenberg Hamiltonian:
\beq
H_H=\sum_{{\bf j},{\bf j}'}\Big[
a_{\u\u}(S_{\bf j}^zS_{{\bf j}'}^z-{1\over4})+
a_{\u\d}(S_{\bf j}^xS_{{\bf j}'}^x+S_{\bf j}^yS_{{\bf j}'}^y)
\Big].
\label{hh}
\eeq
Now the coupling coefficients depend on the number of lattice sites $M$ 
only through $E$:
\[
a_{\u\u}=2t^2 {e^{-2g^2/\omega_0^2}\over\omega_0}
\gamma^*({U+2g^2/\omega_0-E\over\omega_0},-2g^2/\omega_0^2)
\]
\beq
a_{\u\d}=2t^2 {e^{-2g^2/\omega_0^2}\over\omega_0}
\gamma^*({U+2g^2/\omega_0)-E\over\omega_0},2g^2/\omega_0^2).
\label{coupl1}
\eeq

\subsection{Crossover from Weak to Strong Electron-Phonon Coupling}

The energy-dependent coefficients in $H_{eff}$ simplify 
substantially for weak and
strong coupling. Relevant are low energies $E$ which represent
the poles of the resolvent in Eq. (\ref{resolvent0}). To avoid the
poles of the incomplete Gamma function, the following discussion is
restricted to energies 
\beq
E\le U+2g^2/\omega_0.
\label{restr0}
\eeq
It will be shown that in this case there exist poles of the
projected Green's function with $E_j\le0$.
The restriction (\ref{restr0}) implies that the 
approximated diagonal Green's function $G_2$ in Eq. (\ref{g2}) 
has only negative matrix elements:
\[
G_2=(E-2g^2/\omega_0-U-\sum_\bj N_\bj)_2^{-1}<0.
\]
Thus $H_{0,1}G_2H_{0,1}^T$ is a negative matrix and the projected
Green's function
\[
(E-H_{0,1}G_2H_{0,1}^T)_0^{-1}
\]
has all poles $E_j$ on the negative real axis. The relevant parameter
in our truncated continued fraction is $U/\omega_0+2g^2/\omega_0^2\gg1$.
This allows a free tuning of the electron-phonon coupling $g$, as long
as $U$ is sufficiently large.

The weak-coupling limit corresponds with the Hubbard model.
For the latter (i.e. for $g=0$) it is known that a $1/U$ expansion 
at half filling gives in leading
order an isotropic Heisenberg model with
coupling coefficients \cite{fulde} 
\[
a_{\u\u}= a_{\u\d}={2t^2 \over U}.
\]
A similar result was obtained for the Holstein model \cite{freericks93}.
It should be noticed that the energy dependence disappears
in this expansion.
For nonzero but small $g$ ($g/\omega_0\ll 1$) the coupling coefficients
of Eq. (\ref{coupl1}) have the asymptotic behavior 
\[
a_{\u\u}\sim a_{\u\d}\sim{2t^2 \over 2g^2/\omega_0+U-E}.
\]
In the opposite regime, where the electron-phonon coupling
is strong (i.e. $g/\omega_0\gg 1$), the incomplete Gamma function
is approximated by
\[
\gamma^*({U+2g^2/\omega_0-E\over\omega_0},\mp 2g^2/\omega_0^2)
\sim {\omega_0\over U+2g^2/\omega_0-E}
\sum_{m\ge0}{1\over m!}(\pm 2g^2/\omega_0^2)^m 
={\omega_0e^{\pm 2g^2/\omega_0^2}\over U+2g^2/\omega_0-E}.
\]
Thus the coupling coefficients of the pseudospin-1/2 Hamiltonian are
\[
a_{\u\u}\sim{2t^2 \over 2g^2/\omega_0+U-E},\ \ \ \ a_{\u\d}\sim 0.
\]
In this limit the effective Hamiltonian is diagonal and accidentally
degenerate.

The crossover regime, within the restriction of Eq. (\ref{restr0}), is 
shown in Fig. 1.
It indicates that at weak electron-phonon coupling there is a strong
isotropic pseudospin-pseudospin coupling, whereas a strong electron-phonon 
coupling implies a strong pseudospin-pseudospin coupling only for the
$S^z$ component but a weak one for $S^x$ and $S^y$.
Thus the tuning of the electron-phonon interaction is given by
a crossover from an isotropic Heisenberg to an Ising model. This may
be accompanied by a sequence of crossovers and/or phase transitions.

The pole from the groundstate of the projected resolvent $(E-H_H)^{-1}$ 
is easily evaluated in the asymptotic regimes. For the 
weak-coupling regime it is
\[
E_H=g^2/\omega_0+U/2-\sqrt{(g^2/\omega_0+U/2)^2-2t^2\lambda_H}
\]
and for the strong-coupling regime
\[
E_I=g^2/\omega_0+U/2-\sqrt{(g^2/\omega_0+U/2)^2-2t^2\lambda_I},
\]
where $\lambda_H$ ($\lambda_I$) is the lowest eigenvalue of the
isotropic Heisenberg (Ising) Hamiltonian with unit coupling, respectively.

\section{Conclusions}

Starting from a Hamiltonian with short-range Coulomb and $E\otimes\beta$
Jahn-Teller interaction, a system of spinless fermions was studied at
half filling. An effective Hamiltonian $H_{eff}$ was derived under 
the assumption that the kinetic energy (i.e. the hopping term) is always
dominated by the local interaction energy. In the absence of the 
electron-phonon interaction this leads to the well-known isotropic 
pseudospin-1/2 Heisenberg Hamiltonian for $H_{eff}$. A weak
electron-phonon interaction suppresses the pseudospin-1/2 interaction of
the effective Hamiltonian, and an increasing electron-phonon interaction
develops an anisotropy, where the pseudospin-pseudospin interaction in the $xy$
plane decreases like $\exp(-4g^2/\omega_0^2)$ with the electron-phonon
coupling constant $g$ and the pseudospin-pseudospin interaction in the
$z$ direction decreases like $1/g^2$.

\vs
\noindent
Acknowledgement:

\noindent
The author is grateful to K.-H. H\"ock, P. Riseborough, and D. Schneider 
for interesting discussions. 
This work was supported by the Deutsche Forschungsgemeinschaft through
Sonderforschungsbereich 484.

\section*{Appendix A}

\[
a_{\u\u}:=
2t^2 \sum_{m,m'\ge0}{
\langle 0,\u|m,0,0\rangle\langle m,0,0|0,\u\rangle
\langle 0,\d|m',0,0\rangle\langle m',0,0|0,\d\rangle
\over z-[\omega_0(m+m')-(N-2)g^2/\omega_0+U]}
\]
\[
=2t^2 e^{-2g^2/\omega_0^2}\sum_{m,m'\ge0}{1\over m!m'!}
{(g^2/\omega_0)^{m+m'}
\over z-[\omega_0(m+m')-(N-2)g^2/\omega_0+U]}.
\]
The double sum is reduced to a single sum 
\[
=2t^2 e^{-2g^2/\omega_0^2}\sum_{m\ge0}{1\over m!}
{(2 g^2/\omega_0)^m
\over z-\omega_0m+(N-2)g^2/\omega_0-U}
\]
and
\[
a_{\u\d}:=
2t^2 \sum_{m,m'\ge0}{
\langle 0,\u|m,0,0\rangle\langle m,0,0|0,\d\rangle
\langle 0,\d|m',0,0\rangle\langle m',0,0|0,\u\rangle
\over z-[\omega_0(m+m')-(N-2)g^2/\omega_0+U]}
\]
\[
=2t^2 e^{-2g^2/\omega_0^2}\sum_{m,m'\ge0}{1\over m!m'!}
{(-g^2/\omega_0)^{m+m'}
\over z-[\omega_0(m+m')-(N-2)g^2/\omega_0+U]}
\]
The double sum is again reduced to a single sum
\[
=2t^2 e^{-2g^2/\omega_0^2}\sum_{m\ge0}{1\over m!}
{(-2g^2/\omega_0)^m
\over z-\omega_0m+(N-2)g^2/\omega_0-U}.
\]
These expressions are related to the incomplete Gamma function 
\cite{abramowitz}
\[
\gamma^*(a,y)=\sum_{m\ge0}{1\over m!}{(-y)^m\over a+m}
\]
such that
\[
a_{\u\u}=2t^2 {e^{-2g^2/\omega_0^2}\over\omega_0}
\gamma^*({U-z-2(M-1)g^2/\omega_0\over\omega_0},-2g^2/\omega_0^2)
\]
\[
a_{\u\d}=2t^2 {e^{-2g^2/\omega_0^2}\over\omega_0}
\gamma^*({U-z-2(M-1)g^2/\omega_0)\over\omega_0},2g^2/\omega_0^2).
\]

\begin{figure} 
\begin{center}
\includegraphics[scale=0.8]{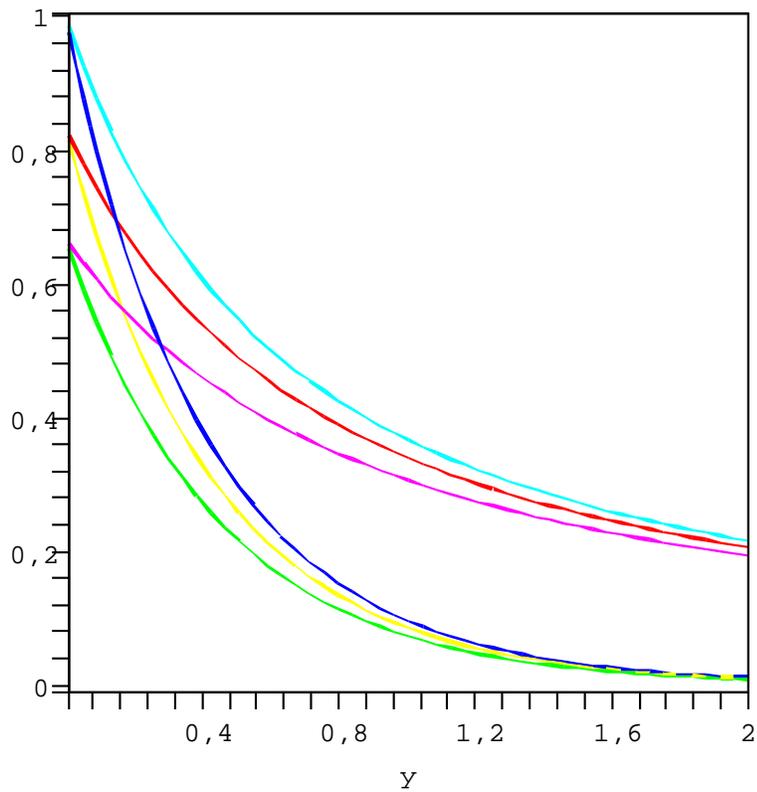}
\end{center} 
\caption{Coupling coefficients of the effective pseudospin-1/2 Hamiltonian
of Eq. (\ref{hh}) as a function of $y=2g^2/\omega_0^2$:
$a_{\u\u}$ (upper set of curves) and $a_{\u\d}$ (lower set of curves) in units of 
$t^2/\omega_0$ for $(U-E)/\omega_0=1.0, 1.2, 1.5$ 
(from top to bottom in each set of curves).}
\end{figure}

\end{document}